\pdfoutput=1
\documentclass[runningheads]{llncs}
\usepackage{graphicx}
\usepackage{mathtools}
\usepackage{microtype}
\usepackage{tikz}
\usetikzlibrary{positioning}
\usepackage[T1]{fontenc}
\usepackage[hidelinks]{hyperref}

\newcommand{\mathcheckurl}{\href{https://uwaterloo.ca/mathcheck}{uwaterloo.ca/mathcheck}}

\begin{document}
\renewcommand{\doi}[1]{\url{https://doi.org/#1}}
\title{Nonexistence Certificates for Ovals in a Projective Plane of Order Ten}
\author{Curtis Bright\inst{1,2} \and
Kevin K. H. Cheung\inst{2} \and
Brett Stevens\inst{2} \and
Ilias Kotsireas\inst{3} \and
Vijay Ganesh\inst{1}}
\authorrunning{C. Bright et al.}
\institute{University of Waterloo, Department of Electrical and Computer Engineering \and
Carleton University, School of Mathematics and Statistics \and
Wilfrid Laurier University, Department of Physics and Computer Science
}
\maketitle
\begin{abstract}
In 1983, a computer search was performed for ovals in a projective plane of order ten.
The search was exhaustive and negative, implying that such ovals do not exist.
However, no nonexistence certificates were produced by this search, and to the best of our
knowledge the search has never been independently verified.
In this paper, we rerun the search for ovals in a projective plane of order ten and produce
a collection of nonexistence certificates that, when taken together, imply that such ovals do not exist.
Our search program uses the cube-and-conquer paradigm from the field of satisfiability (SAT) checking,
coupled with a programmatic SAT solver and the nauty symbolic computation library
for removing symmetries from the search.

\end{abstract}
\setcounter{footnote}{0}

\section{Introduction}

Projective geometry---a generalization of the familiar Euclidean geometry where parallel
lines do not exist---has been
extensively studied since the 1600s.  A special case of projective geometry occurs
when only a finite number of points exist.  A two-dimensional projective
geometry with a finite number of points is known as a \emph{finite projective plane}.

Despite a huge amount of study
some basic questions about finite projective planes are still open---for example,
how many points can a finite projective plane contain?
It is well-known~\cite{hall1955finite} that a finite projective plane must contain
$n^2+n+1$ points for some integer~$n$ (known as the plane's \emph{order})
and finite projective planes can be explicitly constructed in all orders that are prime powers.
The order six case is excluded by a theoretical result of Bruck and Ryser~\cite{bruck1949nonexistence}
making ten the first uncertain order.

In the 1970s and 1980s, a significant amount of mathematical ingenuity
and computer searches successfully eliminated the possibility of a projective plane
of order ten~\cite{lam1991search}.  Today, this remains one of the most prominent achievements
of computational combinatorial classification~\cite{kaski2006classification}.
The search was made feasible due to results
of MacWilliams, Sloane, and Thompson~\cite{macwilliams1973existence}
concerning the error-correcting code generated by a hypothetical projective plane of order ten.
They showed that the weight distribution of this code depends on just two unknown parameters.
One of these parameters is the number of ovals that exist in the projective plane
of order ten---here an \emph{oval} being a set of twelve points,
no three of which are collinear.

In 1983, Lam, Thiel, Swiercz, and McKay~\cite{lam1983nonexistence} showed the
nonexistence of ovals in a projective plane of order ten via a computer search.
The search space is of a significant size and required about 4,400 hours of computation
time on the supermini computer VAX 11/780 (clock speed 5~MHz) to search exhaustively.
Because of the nature of the search, Lam et al.~specifically encouraged an
independent verification:

\begin{quote}
Since the existence of ovals is an important question, we hope that someone will
do an independent search to verify the result.
\end{quote}

Despite this hope, there has been little published work independently verifying the search for ovals
or their subsequent searches~\cite{lam1989non,lam1986nonexistence}
that culminated in the proof that projective planes of order ten do not exist.
In his 2011 master's thesis, Roy~\cite{roy2011confirmation}
performed a verification of the nonexistence of a projective
plane of order ten using about 35,000 hours on a cluster of desktop machines.
However, he did not specifically run a search for the ovals as it was nonessential to his ultimate goal.
To the best of our knowledge, there has been no published work specifically replicating the search for ovals.

In this paper, we report our results on verifying the nonexistence of ovals in a projective
plane of order ten.  Our method relies on a satisfiability (SAT) solver and produces certificates
that a third party can use to verify that our search completed successfully.
In total, our search used about 1,850 core hours on the supercomputer Graham at the University
of Waterloo (clock speed 2.1~GHz) and produced SAT proofs that when compressed use about 3 terabytes of storage.

In addition to a using a SAT solver our method also takes advantage of the nauty symbolic computation
library~\cite{mckay2014practical} to reduce the size of the search space by eliminating
redundant symmetries.  We present the necessary background on projective geometry, satisfiability checking,
and symbolic computation in Section~\ref{sec:prelim},
describe our SAT encoding in Section~\ref{sec:sat},
give details on our implementation and results in Section~\ref{sec:results},
and finally discuss future work in Section~\ref{sec:future}.

\section{Preliminaries}\label{sec:prelim}

The main background necessary to understand our results are some familiarity with
projective geometry (see Section~\ref{subsec:projective}), satisfiability checking
(see Section~\ref{subsec:satisfiability}), and symbolic computation (see Section~\ref{subsec:symbolic}).

\subsection{Projective geometry}\label{subsec:projective}

A \emph{finite projective plane} of order $n$ is a collection of $n^2+n+1$ points
and $n^2+n+1$ lines and an incidence relationship between points and lines where
any two points are incident with a unique line and any two lines
are incident with a unique point.  Furthermore, every line is incident with $n+1$ points
and every point is incident with $n+1$ lines.

An \emph{oval} of a projective plane of even order~$n$ is a set of $n+2$ points
(or $n+1$ points when~$n$ is odd) with no three points collinear (incident with the same line).
It can be shown that it is not possible to find a larger set of points
with no three points collinear~\cite{hall1980configurations}, but no characterization
of ovals in general projective planes is known.  In particular, prior to the search
of Lam et al.~\cite{lam1983nonexistence} it was not known if a
projective plane of order ten could contain ovals or not.

From a computational perspective, a convenient way of representing a finite projective plane
of order~$n$ is by a square $\{0,1\}$ incidence matrix whose $(i,j)$th entry contains a $1$
exactly when the $i$th line is incident to the $j$th point.
We say that two $\{0,1\}$-vectors \emph{intersect} when they share a~$1$ in the same location
and the \emph{weight} of a $\{0,1\}$-vector is the number of nonzero entries it contains.
In this framework, a projective plane of order~$n$ is a $\{0,1\}$-matrix
with $n^2+n+1$ rows that each have weight $n+1$ and pairwise intersect exactly once
(and similarly for the columns).
In other words, a $\{0,1\}$-matrix $A$ with $n^2+n+1$ rows
represents a projective plane exactly when it satisfies
$AA^T = A^TA = nI + J$
where $I$ denotes the identity matrix and~$J$ denotes the matrix consisting of all~$1$s.
Two projective planes that are identical up to row or column permutations are called
\emph{isomorphic} and we call a submatrix of a projective plane a \emph{partial projective plane}.

Suppose that $A$ is a projective plane of order ten that contains an oval.  Without loss
of generality we assume that the first twelve points of the plane consist of an oval.
By definition, each pair of points in the oval must define a unique line, and therefore
there are $\binom{12}{2}=66$ lines incident to the oval.
Without loss of generality, we assume these lines are ordered
in lexicographically increasing order.  In other words, the
first $66$ rows of $A$ have the form
\[ B = \left[\begin{array}{@{\;}c@{\;\;}|@{\;\;}c@{\;}}
\mathtt{110000000000} & \\
\mathtt{101000000000} & \\
\rule{0em}{1.1em}\smash\vdots & \smash{\raisebox{0.75em}{$B'$}} \\
\mathtt{000000000011} &
\end{array}\right] .
\]
The first twelve columns contain two $1$s on each row, so
$B'$ must contain nine~$1$s in each row.
Furthermore, by definition of a projective plane
each column in~$B'$ must intersect each of the first
twelve columns.
Each $1$ in $B'$ induces an intersection with
two of the first twelve columns, so each column in $B'$ contains exactly six~$1$s.

\begin{figure}[t]
\input{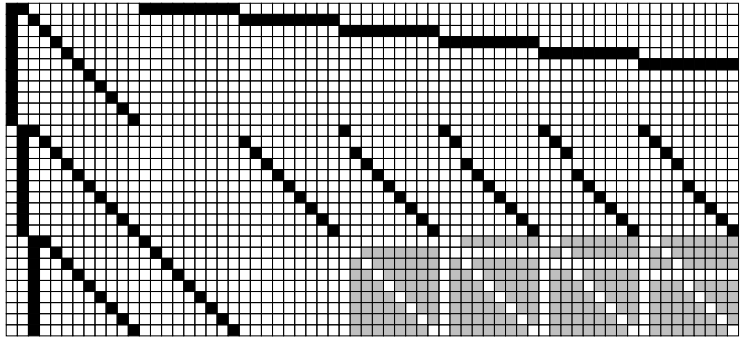}
\caption{The upper-left $30\times66$ submatrix of~$B$ under the assumption
that the rows are lexicographically ordered
and the columns outside the oval are lexicographically ordered.
Black entries denote~$1$s, white entries denote~$0$s, and gray entries
are unknown.}\label{fig:initial}
\end{figure}

Without loss of generality, we assume the columns of~$B'$ are sorted in lexicographic order. 
This implies the first nine columns of~$B'$ will be incident with the first line (the line through
the first and second points).  As noted by~\cite{lam1982feasibility},
this also means the $i$th column of $B'$ (for $1\leq i\leq 9$)
will be incident with the line through the third and $(3+i)$th points.
We call the nine columns of $B'$ that are incident with the $i$th row the $i$th \emph{block}.
In general, all blocks' columns may be ordered
similarly to those in the first block~\cite{lam1982feasibility},
and this fixes the first two~$1$s in each column of $B'$.
Figure~\ref{fig:initial} contains a visual depiction of the first~30 rows
of~$B$ up to the sixth block.

Some entries of $B'$ are still undetermined (shown as gray in Figure~\ref{fig:initial}).
At this stage, it is still uncertain if they can be completed in a consistent way to make
$B'$ a partial projective plane---since the above description
assumes that an oval exists in~$A$.  Thus, a proof that there is
no way of completing the unknown entries of $B'$ in a consistent way would
also imply the nonexistence of ovals in a projective plane of order ten.

The \emph{symmetry group} of a matrix is the group of row and column permutations
that fix the entries of the matrix.
For example, consider the symmetry group~$S$ of the first twelve columns of~$B$.
Each row of this submatrix is completely specified by the two columns incident to it,
so any column permutation completely specifies a row permutation that
undoes the permutation.  It follows that~$S$
is isomorphic to~$S_{12}$, the symmetric group on twelve elements.

The group~$S$ acts on the entries of~$B'$ as follows:
Given a permutation $\varphi\in S$ the row permutations from $\varphi$ are applied to
the entries of~$B'$, then column permutations are applied to reorder
its columns in lexicographic order.
The result $\varphi(B')$ is a partial projective
plane that is isomorphic to~$B'$.
To avoid duplication of work,
any search for $B'$ should ideally avoid exploring parts of the search
space that are isomorphic under~$S$.  Exploiting this leads to a huge
reduction in the size of the search space, since~$S$
contains about 479 million permutations.

\subsection{Satisfiability checking}\label{subsec:satisfiability}

Given a formula of Boolean logic, satisfiability (SAT) checking is to determine
whether or not the formula is \emph{satisfiable}---that is, is there a way of assigning
true and false to its variables that results in the whole formula becoming true? 
A \emph{SAT solver} is a program that performs SAT checking on a given formula.
Modern SAT solvers require their input to be given in conjunctive normal form or CNF:
if $x$ is a Boolean variable then $x$ and $\lnot x$ are known as \emph{literals},
expressions of the form $l_1\lor\dotsb\lor l_n$ for literals $l_i$ are known as \emph{clauses},
and expressions of the form $c_1\land\dotsb\land c_m$ for clauses $c_i$ are in \emph{CNF}.
The literal $x$ is satisfied when $x$ is assigned true, $\lnot x$ is satisfied when
$x$ is assigned false, $l_i\lor\dotsb\lor l_n$ is satisfied when at least one $l_i$ is satisfied,
and $c_1\land\dotsb\land c_m$ is satisfied when every $c_i$ is satisfied.

In order to reduce the search for ovals in a projective plane of order ten to a SAT problem
we use the incidence structure described in Section~\ref{subsec:projective} that was based
on the assumption that ovals exist.  The SAT instance will have a solution
when there is a completion of the unknown entries of the
matrix~$B$
to a partial projective plane---so showing the instance has no solution implies that
no ovals exist.

SAT solvers are effective as combinatorial search tools---%
for example, they were used to resolve the first open case
of the Erd\H{o}s discrepancy conjecture~\cite{konev2015computer}.
The \emph{cube-and-conquer} SAT solving paradigm has been particularly effective
at solving very large combinatorial search problems~\cite{heule2017solving}.  First developed
by Heule, Kullmann, Wieringa, and Biere for computing van der Waerden numbers~\cite{heule2011cube},
the cube-and-conquer method
has since been used to resolve the Boolean Pythagorean triples problem~\cite{heule2016solving}
and determine the value of the fifth Schur number~\cite{heule2018schur}.

A \emph{cube} is a formula of the form $l_1\land\dotsb\land l_n$ where $l_i$ are literals.
In the cube-and-conquer paradigm a SAT instance is split into a
number of distinct subinstances specified by cubes.  Each subinstance contains a
single cube and the cube is assumed to be true for the purposes of solving the subinstance.
The cubes are typically generated by running a ``cubing solver'' on the SAT instance
which attempts to find a set of cubes which split the instance into subinstances of
approximately equal difficulty.
After the cubes have been generated a ``conquering solver''
solves the subinstances either in sequence or in parallel.
Ideally, the literals in each cube are added to the solver as incremental assumptions.
In this case, after each cube is solved the assumptions
are removed and the literals from the next cube are
added without restarting the SAT solver.

\subsection{Symbolic computation and SAT+CAS}\label{subsec:symbolic}

Symbolic computation is a branch computer science devoted to manipulating and
simplifying mathematical expressions.  Many computer algebra systems
(CASs) are available today that contain extensive symbolic computation functionality
from a huge number of mathematical domains.  However, although CASs contain many sophisticated
algorithms, they have not typically been optimized to perform searches in the way that
SAT solvers have~\cite{abraham2015building,bright2019sat}.

For problems that need \emph{both} mathematical sophistication
and finely-tuned search it can be useful to combine
computer algebra and SAT solvers~\cite{davenport2020symbolic}.
Recently SAT+CAS methods have been used in a number of various problems---for example,
they have been used to verify the correctness of Boolean arithmetic circuits~\cite{kaufmann2019verifying},
improve the best known
result in the Hadwiger--Nelson plane-colouring problem~\cite{heule2019trimming},
find many new algorithms for multiplying $3\times3$ matrices~\cite{heule2019new},
and improve the best known result in the Ruskey--Savage hypercube conjecture~\cite{zulkoski2017combining}.

In addition to a SAT solver
we use the nauty symbolic computation library~\cite{mckay2014practical}
in order to show the nonexistence of ovals in a projective plane of order ten.
We call nauty from within the callback function of a ``programmatic'' SAT solver.
A solver is called \emph{programmatic} if it allows learning clauses on-the-fly
through a piece of code supplied to the SAT solver.
A programmatic SAT solver will run the supplied code from time to time
as it is performing its search.
The code will examine the current assignment to the variables and test whether
the current assignment may be discarded (possibly using knowledge queried from a CAS).
If the assignment can be discarded a clause is added to the SAT instance on-the-fly
that blocks the current assignment (and ideally other similar assignments).
Programmatic SAT solvers were introduced by Ganesh et al.~\cite{ganesh2012lynx} in order
to solve an RNA folding problem.  They have since been used
to search for various combinatorial objects such as Williamson matrices~\cite{bright2020applying},
best matrices~\cite{bright2019satcas}, and complex Golay sequences~\cite{bright2019complex}.

\section{Satisfiability encoding}\label{sec:sat}

We now describe the encoding that we use to search for ovals in a projective plane of order ten.
As described in Section~\ref{subsec:projective},
we may assume a number of entries of this projective plane have been fixed in advance,
including all entries in the first twelve columns and all entries in the first 21 rows
(see Figure~\ref{fig:initial}).
Specifying these entries removes a substantial amount of symmetry from the search space,
however, as described in Section~\ref{subsec:projective}, the remaining search space is still
symmetric under the action of the group~$S$ generated by permuting the twelve points
of the oval.
In Section~\ref{subsec:encoding},
we give our basic encoding without removing symmetries from~$S$.
In Section~\ref{subsec:breaking},
we provide a programmatic SAT method of removing symmetries from the group~$S$.

\subsection{Basic SAT encoding}\label{subsec:encoding}

Following Section~\ref{subsec:projective}, let $B$ be the first 66 rows
of the incidence matrix of a partial projective plane of order ten whose
first twelve points form an oval.  As previously outlined, up to isomorphism some points
of~$B$ can be assumed in advance, but most points remain unspecified.  For each
unspecified point we define a Boolean variable $b_{i,j}$ that will
be true exactly when the $i$th line is incident to the $j$th point, i.e.,
the $(i,j)$th entry of~$B$ is~$1$.

We now give properties that necessarily hold in~$B$
as Boolean constraints in conjunctive normal form.  In particular, we encode the two facts
that (1) columns of $B$ intersect at most once and (2) each column of $B$ intersects
a column in the oval at least once.
A similar encoding has been previously used to
verify MacWilliams et al.'s result that vectors of weight 15 do
not exist in the rowspace of any projective plane of order ten~\cite{bright2019nonexistence}.

\subsubsection{Columns intersect at most once}

Let $i$ and $j$ be arbitrary column indices of~$B$, so
$i,j\in\{1,\dotsc,111\}$.  By definition of a projective plane
these columns cannot intersect twice, so we know there do not exist
rows $k$ and $l$ mutually incident to columns $i$ and $j$.  In Boolean logic
we write this constraint as
\[ \bigwedge_{1\leq k<l\leq 66} (\lnot b_{k,i}\lor\lnot b_{k,j}\lor\lnot b_{l,i}\lor\lnot b_{l,j}) . \]

\subsubsection{Each column intersects a column in the oval}

Let $i$ be an arbitrary column in the oval, i.e., $i\in\{1,\dotsc,12\}$ and let $j$
be an arbitrary column not in the oval, i.e., $j\in\{13,\dotsc,111\}$.  By definition
of a projective plane columns~$i$ and~$j$ must intersect somewhere in the plane.
Since all rows incident with column~$i$ occur in the first $66$ rows,
the intersection of columns~$i$ and~$j$ must occur in~$B$.
In Boolean logic we write this constraint as
$\bigvee_{k:B[k,i]=1}b_{k,j}$.

\subsubsection{Abbreviated constraints}

For the first set of constraints
there are $\binom{111}{2}\cdot\binom{66}{2}\approx\text{13 million}$ clauses of the first form
and $12\cdot99=1188$ clauses of the second form.
After removing variables whose values are already fixed there are 2696 undetermined variables
in these clauses.

SAT solvers often perform better if the number of constraints can be significantly decreased.
In our case, we found that it was only necessary to consider a submatrix of~$B$
before reaching a contradiction.  In particular, our primary searches only used the variables
in the blocks 2 to 6 (columns 22 to 66).  The first block was skipped since its columns
did not intersect the columns of any other block in the known entries (see Figure~\ref{fig:initial}).
This increases the efficiency of the search because contradictions are generally easier to derive
from two already-intersecting columns.
Using columns 22 to 66 meant there were $\binom{57}{2}\cdot\binom{66}{2}\approx\text{3.4 million}$ clauses
of the first form, $12\cdot45=540$ clauses of the second form, and 1199 unknown variables.

\subsubsection{Known row intersections}

We included one further set of constraints that, while not strictly necessary, improved the
performance of the SAT solver by enforcing row intersections that must occur.
In particular, note that rows 2--6 must intersect rows 22--66 in~$B$ and all $1$s in
row $i\in\{2,\dotsc,6\}$ outside the oval occur in the columns $B_i\coloneqq\{4+9i,\dotsc,12+9i\}$.
Thus, we also included clauses of the form
$\bigvee_{k\in B_i} b_{j,k}$
for rows $i\in\{2,\dotsc,6\}$ and $j\in\{22,\dotsc,66\}$ that do not intersect in the oval.

\subsection{Symmetry breaking}\label{subsec:breaking}

The encoding described in Section~\ref{subsec:encoding} could in theory be used to show there is no
way of completing the unknown entries of~$B$ subject to the given constraints.  However, as
discussed in Section~\ref{subsec:projective} the
search space is symmetric under the action of relabelling the twelve points of the oval (while
appropriately reordering the rows and remaining columns to preserve our lexicographic presentation
of the search space).
Since this is an enormous group of symmetries it is worthwhile developing a method that will
reduce or ``break'' these symmetries.
Using an ``orderly generation'' algorithm~\cite{read1978every,royle1998orderly} is one
way to avoid generating isomorphic partial solutions at each stage of the search.
Our approach is similar, though it will only avoid isomorphic partial solutions
violating a property that we can show the entries of~$B$ satisfy (up to isomorphism).

Mathon~\cite{mathon1981partial} provided a characterization of an oval in a projective plane
of order ten in terms of $K_{12}$, the complete graph on vertices $\{1,\dotsc,12\}$.  Note that a \emph{1-factor}
of a graph is a perfect matching of its edges and a \emph{1-factorization} of a graph
is a decomposition of its edges into 1-factors.  If rows denote edges and columns denote points then
the first twelve columns of~$B$ are precisely the incidence matrix of $K_{12}$.
Every column of~$B$ outside the oval contains six~$1$s on rows that will not be adjacent
(as edges of $K_{12}$).  Therefore, each column of~$B$ outside the oval forms a $1$-factor of $K_{12}$.

Furthermore, consider the set of columns in the first block of~$B$.  These rows are all incident
to the row through points~$1$ and~$2$.  The other five~$1$s in each column must
each occur on distinct rows and will cover the remaining $\binom{10}{2}=45$ rows
through the points $\{3,\dotsc,12\}$.  Therefore, the first
block of~$B$ forms a $1$-factorization of $K_{12}\setminus\{1,2\}\cong K_{10}$ and in general the~$i$th block
forms a $1$-factorization of $K_{12}\setminus\{1,i+1\}$.
Gelling~\cite{gelling1973} determined that there are exactly 396 nonisomorphic 1-factorizations
of~$K_{10}$ and we assume that each has been given a distinct label in the set $\{1,\dotsc,396\}$.

Note that the symmetry group~$S$ generated by permuting the columns of the oval acts
transitively on the set of blocks: there is a permutation in~$S$ that will send any block
to any other block.  Suppose we tag each completed block of $B$ with the label
(as described above) of the 1-factorization that it is isomorphic with.
We may assume that block~2 of $B$ has the minimal label amongst
the blocks of~$B$---if it didn't, we could send the block with minimal label to
block~2 by an appropriate permutation of~$S$.
Our symmetry breaking method will enforce the condition that block~2 has the
minimal label amongst the other blocks of~$B$ for which we are searching (blocks 2--6).  However,
it is not very easy to concisely express this constraint as clauses in Boolean logic.  Therefore, we
make use of the programmatic SAT paradigm in order to enforce this constraint on-the-fly.

\subsubsection{Programmatic symmetry breaking}\label{subsec:programmatic}

A programmatic SAT solver is compiled with a ``callback'' function that often
examines the solver's current assignment (the mapping from variables to truth values).
When the callback function determines the
current state should be discarded it will add clauses to the SAT instance that block
the current assignment.

If all the variables in the~$i$th block have been assigned
and $p$ is one of these variables then we let~$B_i\models p$ denote that variable $p$
has been assigned true.
Suppose all the variables in block~2 and
block $i\in\{3,\dotsc,6\}$
have been assigned.  If the label of block~2 is strictly larger than the label
of block~$i$ we want to block this configuration from the search space.  In such a
case we want to add the Boolean constraint
\[ \bigwedge_{B_2\models p}p \mathrel{\to} \Bigr(\lnot\bigwedge_{B_i\models p} p \Bigl)
\qquad \text{or equivalently} \qquad
\bigvee_{B_2\models p}\lnot p \lor\bigvee_{B_i\models p}\lnot p \]
which says that the $i$th block cannot be assigned the way it currently is
while the second block is assigned the way it currently is.

\section{Implementation and results}\label{sec:results}

Our SAT encoding is implemented as a part of the MathCheck project; our
scripts are open source and available online at \mathcheckurl.  The search proceeds
in three main parts: First, we verify the result of Gelling~\cite{gelling1973} that
there are exactly 396 nonisomorphic 1-factorizations of $K_{10}$.  Second, we generate
396 separate SAT instances (one for each nonisomorphic way of filling in block~2
of~$B$).  The cube-and-conquer method is used in parallel to solve each SAT instance.
A cubing solver generates a set of cubes from each SAT instance and a programmatic
conquering solver is
used to show that (up to the symmetry breaking method of Section~\ref{subsec:programmatic})
there are
58 ways of completing the blocks~2--6.
Finally, we generate a new SAT instance for each of the 58 solutions and verify that
there are no consistent ways of extending these completions to block~7.
Additionally, to increase the confidence that the SAT instances were successfully solved
the SAT solvers produced DRUP (delete reverse unit propagation) certificates~\cite{heule2013trimming}
which were subsequently verified.
A flowchart of these steps is available in this paper's appendix at \mathcheckurl.

\subsection{Generating the SAT instances}

The SAT instances are generated by a Python script that writes the clauses described in
Section~\ref{subsec:encoding} to a file in DIMACS
(Discrete Mathematics and Theoretical Computer Science) CNF format.
The script accepts as a parameter the columns to include in the SAT instance
and by default uses the columns in blocks~2--6 (those used in our primary search).

\subsection{Generating the nonisomorphic 1-factorizations}\label{subsec:1factorizations}

The 396 nonisomorphic 1-factorizations of $K_{10}$ as reported by Gelling~\cite{gelling1973}
can be quickly generated using a straightforward search, but we used a SAT approach
as that was convenient for our purpose.  The SAT instance
only uses the variables in block~2, the columns
of this block corresponding with 1-factors of $K_{12}\setminus\{1,3\}$.
As noted by Gelling, up to
isomorphism the entries of the first 1-factor can be completely assumed.
By the lexicographic ordering assumption the first 1-factor
includes the edge $(2,4)$ and
by permuting the columns $\{5,\dotsc,12\}$ of the oval we can assume the first 1-factor
contains the edges $(5,6)$, $(7,8)$, $(9,10)$, and $(11,12)$.
Gelling also noted that after fixing the first 1-factor there are exactly two ways (up to isomorphism)
of fixing the second 1-factor
and the union of the first two 1-factors either form a 4-cycle and a 6-cycle
or a 10-cycle.

The entries that can be fixed are given to the SAT solver as unit clauses and a programmatic
implementation finds all nonisomorphic 1-factorizations.
Whenever a completion of block~2 is found, the program Traces from the nauty
graph isomorphism library determines if the completion
is new or isomorphic to a previously found completion.  (The graph provided to Traces is
the \emph{incidence graph} representation~\cite{godsil2013algebraic} of the first 12 columns of~$B$ and the columns
of block~2.)  A new completion is recorded for later use and a clause that
blocks the completion (i.e., $\bigvee_{B_2\models p}\lnot p$) is added to the SAT instance
until all possible completions have been examined.  A programmatic implementation
of MapleSAT~\cite{liang2017empirical} confirms the result of Gelling
that 396 nonisomorphic 1-factorizations of $K_{10}$ exist in about 8 seconds.

\subsection{Solving the SAT instances: Cubing}

We now generate 396 distinct SAT instances, one for each of the
396 nonisomorphic ways of filling in block~2.  Variables from blocks 2--6
are used in each SAT instance, with the variables in block~2
completely determined by the specific nonisomorphic 1-factorization chosen in each case.
We simplify these instances with the preprocessor of the SAT solver Lingeling~\cite{biere2018lingeling}
which produces proofs of simplification without renaming variables.
After simplification, these instances each contained 912 unknown variables and on average contained
22,883 clauses.  Simplifying all 396 SAT instances requires about 15 minutes in total.

Next, we apply the cubing solver March\_cu~\cite{heule2011cube} on each of the
396 individual SAT instances.  The conquering solver (see Section~\ref{subsec:conquering})
typically performs better when the variables in the cubes are not split across blocks.
Thus, we modified March\_cu so that it only produces cubes using variables occurring in the same block as the
first variable in the cubes.  We controlled the cubing cutoff using the \mbox{\kern1sp\texttt{-n}}
parameter of March\_cu which stops cubing once the number of free variables falls
below the given bound.  Each block contains 228 unknown variables and we stop cubing
once the subproblems specified by each cube contain at least 228 fewer free variables
than the original instance.
On average, March\_cu produced about 180,000 cubes per SAT instance and spent about
17.5 total hours in this step.

\subsection{Solving the SAT instances: Conquering}\label{subsec:conquering}

The majority of the search work was done by the conquering solver.  A programmatic
version of MapleSAT~\cite{liang2017empirical} was used to complete this step.
Each of the 396 SAT instances along with the cubes previously computed for each instance
were given to separate instances of MapleSAT and solved in parallel.  The literals in
each cube were specified as incremental assumptions~\cite{nadel2012efficient}, so that
it was not necessary to restart the SAT solver after solving each cube.

The programmatic encoding from Section~\ref{subsec:programmatic} was used to ignore
any completions of blocks 3--6 whose label was strictly smaller than the label of
block~2 (which was fixed in each SAT instance).  The label of each block completion
can be computed by calling nauty on the incidence graph representation of the block
(as described in Section~\ref{subsec:1factorizations}).  However, these incidence graphs
contain 87 vertices (from 66 rows and 21 columns) and we found there was significant
overhead from calling nauty in this way.

Our final implementation makes use of a simpler check based on Gelling's observation
that up to isomorphism each pair of columns in a block are of two types
(either a 4-cycle and 6-cycle or a 10-cycle).  Given a complete block,
we check all $\binom{9}{2}=36$ pairs of columns and generate the
\emph{cycle pattern} for each block up to isomorphism---for example,
one cycle pattern consists of the case when all pairs of columns form 10-cycles.
In general, a cycle pattern graph on 9 vertices is constructed where two vertices
are adjacent exactly when their associated two columns form a 10-cycle.
Using nauty we determined that the 396 distinct block types gave rise to 359
distinct cycle patterns and
in our programmatic implementation we used the cycle
pattern as a proxy for determining the block label.  In most cases the cycle pattern
could be used to uniquely identify the block label, but otherwise the block label
was assigned to the largest possible label consistent with the given cycle pattern (i.e., the most
pessimistic choice in terms of symmetry breaking).

Following~\cite{lam1986nonexistence}, block labels were chosen for the blocks by
sorting the blocks in ascending order by the size of their stabilizer groups.  Additionally,
blocks with identical cycle patterns were given adjacent labels when possible
in order to minimize the impact of the above pessimistic choice.

In total, this step required about 1,832 core hours on a cluster of
Intel E5-2683 CPUs running at 2.1 GHz.  The search produced 58 valid
completions of the blocks 2--6 (see \mathcheckurl\ for one explicit completion).
Whenever a valid completion~$B$
was found, a clause $\bigvee_{B\models p}\lnot p$
was programmatically added to the SAT instance.
The added clause blocked the completion from occurring again later in the search.

Finally, for each of the 58 completions of blocks 2--6 a SAT instance was generated that included the
constraints from blocks~2--7 and a cube specifying the completion (i.e., $\bigwedge_{B\models p}p$).
It was found that none of the completions of blocks~2--6 could be extended to block~7
and this final step required less than a second.

\subsection{Certificate verification}

The runs from the solvers produced DRUP proofs totalling about 33 terabytes.
These were verified using the proof verification tool DRAT-trim~\cite{wetzler2014drat}
which was also used to trim and compress the proofs.  These optimized proofs were
archived using 7z data compression and produced archives totalling about~3~terabytes.
These archives are available from the authors by request.

In order for the proofs to be verified by DRAT-trim the clauses which were programmatically generated
during the solver's run also need to be provided to DRAT-trim.  One way of doing this is to
add the programmatic clauses directly into the CNF file provided to DRAT-trim.  However, this
method was found to suffer from very poor performance because this significantly
increased the size of the initial active clause database tracked by DRAT-trim.

To get around this we modified
DRAT-trim to support the addition of ``trusted'' clauses midway through the proof.  Normally, each
step of a proof consists of either an \emph{addition} or \emph{deletion} to the active clause database.
In the case of an addition, DRAT-trim verifies that the added clause is a logical consequence%
\footnote{DRAT-trim also supports a more general kind of provability termed
``resolution asymmetric tautology'' that we did not use in our proofs.}
of the current set of active clauses.
In our proofs we have a third kind of step, a \emph{trusted addition} that adds the clause
into the current set of active clauses without checking its provability.  The justification
for these clauses relies on our symmetry breaking method and not on a property
easily checkable in Boolean logic, so the symmetry breaking clauses were not verified by DRAT-trim.
However, if you believe in the correctness of our SAT encoding, generation scripts, DRAT-trim,
and the trusted additions (whose correctness relies on our symmetry breaking method and a call to nauty) then you
must believe in our certificates.

These proofs were checked using a system configured to limit each core to at most 4~GB of memory.
In order to meet this limit it was necessary to ensure that each proof did not grow too large.
To do this, March\_cu was used generate a second ``toplevel'' set of cubes that
partitioned the $i$th SAT instance into $398-i$ subinstances.
(As the label increased fewer subinstances were used because the instances
became easier due to symmetry breaking.)  Each of the subinstances were solved and had their proofs
verified separately (each using at most 4~GB of memory and~10 minutes of computing time).

\section{Conclusions and future work}\label{sec:future}

In this paper we have completed an independent search showing the nonexistence of ovals
in a projective plane of order ten.  This was accomplished using a reduction to the Boolean
satisfiability problem along with a SAT solver to show the resulting SAT instances are unsatisfiable.
However, in order to make the amount of computation feasible
it was necessary
to use a symmetry breaking method.  We used a ``programmatic'' SAT solver coupled with
the symbolic computation library nauty~\cite{mckay2014practical} in order to learn symmetry breaking clauses on-the-fly
during the search.

Our implementation uses the SAT+CAS interface as developed by the MathCheck
project~\cite{zulkoski2017combining}.  We are currently working on using MathCheck to
verify more of the searches that were necessary in order to show the nonexistence of projective
planes of order ten~\cite{bright2019nonexistence}.  To date, we have verified the searches
of MacWilliams et~al.~\cite{macwilliams1973existence}, Carter~\cite{carter1974existence},
and Lam et~al.~\cite{lam1986nonexistence} that show that the rowspace of a
projective plane of order ten does not contain vectors of weight~15 or~16.
A consequence of these searches is that the
weight enumerator of the error-correcting code generated by a projective plane of order
ten can be specified exactly~\cite{macwilliams1973existence}.

In particular, the rowspace of a projective plane of order ten
must contain exactly 24,675 vectors of weight~19.
The search for such vectors is the only case that remains in order to provide a complete SAT-based
independent verification of the nonexistence of projective planes of order ten.
We are currently exploring the feasibility of this and believe MathCheck will be useful in this case
as well.
The same basic encoding can be used but it seems necessary to tailor
the symmetry breaking method and the structure of the search.
This will be the subject of future research.

\paragraph{Acknowledgements}

The authors would like to thank the reviewers and the coordinators at Springer Nature whose work
improved the quality and correctness of this publication.

\bibliographystyle{splncs04}
\bibliography{iwoca}

\begin{thebibliography}{10}
\providecommand{\url}[1]{\texttt{#1}}
\providecommand{\urlprefix}{URL }
\providecommand{\doi}[1]{https://doi.org/#1}

\bibitem{abraham2015building}
{\'A}brah{\'a}m, E.: Building bridges between symbolic computation and
  satisfiability checking. In: Proceedings of the 2015 ACM on International
  Symposium on Symbolic and Algebraic Computation. pp.~1--6. ACM (2015).
  \doi{10.1145/2755996.2756636}

\bibitem{biere2018lingeling}
Biere, A.: {CaDiCaL}, {Lingeling}, {Plingeling}, {Treengeling} and {YalSAT}
  entering the {SAT} competition. Proceedings of SAT Competition 2018: Solver
  and Benchmark Descriptions  (2018), \url{http://fmv.jku.at/lingeling}

\bibitem{bright2019nonexistence}
Bright, C., Cheung, K., Stevens, B., Roy, D., Kotsireas, I., Ganesh, V.: A
  nonexistence certificate for projective planes of order ten with weight 15
  codewords. Applicable Algebra in Engineering, Communication and Computing
  (2020). \doi{10.1007/s00200-020-00426-y}

\bibitem{bright2019sat}
Bright, C., Kotsireas, I., Ganesh, V.: {SAT} solvers and computer algebra
  systems: A powerful combination for mathematics. In: Proceedings of the 29th
  Annual International Conference on Computer Science and Software Engineering.
  pp. 323--328. IBM Corp. (2019),
  \url{https://dl.acm.org/doi/10.5555/3370272.3370309}

\bibitem{bright2020applying}
Bright, C., Kotsireas, I., Ganesh, V.: Applying computer algebra systems with
  {SAT} solvers to the {Williamson} conjecture. Journal of Symbolic Computation
   \textbf{100},  187--209 (2020). \doi{10.1016/j.jsc.2019.07.024}

\bibitem{bright2019complex}
Bright, C., Kotsireas, I., Heinle, A., Ganesh, V.: Complex golay pairs up to
  length 28: A search via computer algebra and programmatic {SAT}. Journal of
  Symbolic Computation  (2019). \doi{10.1016/j.jsc.2019.10.013}

\bibitem{bright2019satcas}
Bright, C., {\DJ}okovi{\'c}, D.{\v{Z}}., Kotsireas, I., Ganesh, V.: The
  {SAT+CAS} method for combinatorial search with applications to best matrices.
  Annals of Mathematics and Artificial Intelligence  \textbf{87}(4),  321--342
  (2019). \doi{10.1007/s10472-019-09681-3}

\bibitem{bruck1949nonexistence}
Bruck, R.H., Ryser, H.J.: The nonexistence of certain finite projective planes.
  Canadian Journal of Mathematics  \textbf{1}(1),  88--93 (1949).
  \doi{10.4153/CJM-1949-009-2}

\bibitem{carter1974existence}
Carter, J.L.: On the existence of a projective plane of order ten. Ph.D.
  thesis, University of California, Berkeley (1974),
  \url{https://hdl.handle.net/2027/uc1.c3475138}

\bibitem{davenport2020symbolic}
Davenport, J.H., England, M., Griggio, A., Sturm, T., Tinelli, C.: Symbolic
  computation and satisfiability checking. Journal of Symbolic Computation
  \textbf{100},  1--10 (2020). \doi{10.1016/j.jsc.2019.07.017}

\bibitem{ganesh2012lynx}
Ganesh, V., O'Donnell, C.W., Soos, M., Devadas, S., Rinard, M.C., Solar-Lezama,
  A.: Lynx: A programmatic {SAT} solver for the {RNA}-folding problem. In:
  International Conference on Theory and Applications of Satisfiability
  Testing. pp. 143--156. Springer (2012). \doi{10.1007/978-3-642-31612-8\_12}

\bibitem{gelling1973}
Gelling, E.N.: On 1-factorizations of the complete graph and the relationship
  to round robin schedules. Master's thesis, University of Victoria (1973),
  \url{http://hdl.handle.net/1828/7341}

\bibitem{godsil2013algebraic}
Godsil, C., Royle, G.F.: Algebraic graph theory, vol.~207. Springer Science \&
  Business Media (2013). \doi{10.1007/978-1-4613-0163-9}

\bibitem{hall1955finite}
{Hall Jr.}, M.: Finite projective planes. The American Mathematical Monthly
  \textbf{62}(7P2),  18--24 (1955). \doi{10.2307/2308176}

\bibitem{hall1980configurations}
Hall~Jr., M.: Configurations in a plane of order ten. In: Annals of Discrete
  Mathematics, vol.~6, pp. 157--174. Elsevier (1980).
  \doi{10.1016/S0167-5060(08)70701-5}

\bibitem{heule2018schur}
Heule, M.J.H.: Schur number five. In: Proceedings of the Thirty-Second AAAI
  Conference on Artificial Intelligence. pp. 6598--6606. AAAI Press (2018),
  \url{https://www.aaai.org/ocs/index.php/AAAI/AAAI18/paper/view/16952}

\bibitem{heule2019trimming}
Heule, M.J.H.: Trimming graphs using clausal proof optimization. In:
  International Conference on Principles and Practice of Constraint
  Programming. pp. 251--267. Springer (2019).
  \doi{10.1007/978-3-030-30048-7\_15}

\bibitem{heule2013trimming}
Heule, M.J.H., {Hunt Jr.}, W.A., Wetzler, N.: Trimming while checking clausal
  proofs. In: 2013 Formal Methods in Computer-Aided Design. pp. 181--188. IEEE
  (2013). \doi{10.1109/FMCAD.2013.6679408}

\bibitem{heule2019new}
Heule, M.J.H., Kauers, M., Seidl, M.: New ways to multiply $3\times3$-matrices.
  arXiv preprint arXiv:1905.10192  (2019),
  \url{https://arxiv.org/abs/1905.10192}

\bibitem{heule2016solving}
Heule, M.J.H., Kullmann, O., Marek, V.W.: Solving and verifying the {Boolean}
  {Pythagorean} triples problem via cube-and-conquer. In: International
  Conference on Theory and Applications of Satisfiability Testing. pp.
  228--245. Springer (2016). \doi{10.1007/978-3-319-40970-2\_15}

\bibitem{heule2017solving}
Heule, M.J.H., Kullmann, O., Marek, V.W.: Solving very hard problems:
  Cube-and-conquer, a hybrid {SAT} solving method. In: Proceedings of the
  Twenty-Sixth International Joint Conference on Artificial Intelligence,
  {IJCAI-17}. pp. 4864--4868 (2017). \doi{10.24963/ijcai.2017/683}

\bibitem{heule2011cube}
Heule, M.J.H., Kullmann, O., Wieringa, S., Biere, A.: Cube and conquer: Guiding
  {CDCL} {SAT} solvers by lookaheads. In: Haifa Verification Conference. pp.
  50--65. Springer (2011). \doi{10.1007/978-3-642-34188-5\_8}

\bibitem{kaski2006classification}
Kaski, P., {\"O}sterg{\aa}rd, P.R.J.: Classification Algorithms for Codes and
  Designs. Springer (2006). \doi{10.1007/3-540-28991-7}

\bibitem{kaufmann2019verifying}
Kaufmann, D., Biere, A., Kauers, M.: Verifying large multipliers by combining
  {SAT} and computer algebra. In: 2019 Formal Methods in Computer Aided Design
  (FMCAD). pp. 28--36. IEEE (2019). \doi{10.23919/FMCAD.2019.8894250}

\bibitem{konev2015computer}
Konev, B., Lisitsa, A.: Computer-aided proof of {Erd\H{o}s} discrepancy
  properties. Artificial Intelligence  \textbf{224},  103--118 (2015).
  \doi{10.1016/j.artint.2015.03.004}

\bibitem{lam1982feasibility}
Lam, C., Thiel, L., Swiercz, S.: A feasibility study of a search for ovals in a
  projective plane of order 10. In: Combinatorial Mathematics IX, pp. 349--352.
  Springer (1982). \doi{10.1007/BFb0061988}

\bibitem{lam1991search}
Lam, C.W.H.: The search for a finite projective plane of order 10. The American
  Mathematical Monthly  \textbf{98}(4),  305--318 (1991).
  \doi{10.1080/00029890.1991.12000759}

\bibitem{lam1989non}
Lam, C.W.H., Thiel, L., Swiercz, S.: The non-existence of finite projective
  planes of order 10. Canadian Journal of Mathematics  \textbf{41}(6),
  1117--1123 (1989). \doi{10.4153/CJM-1989-049-4}

\bibitem{lam1986nonexistence}
Lam, C.W.H., Thiel, L., Swiercz, S.: The nonexistence of code words of weight
  16 in a projective plane of order 10. Journal of Combinatorial Theory, Series
  A  \textbf{42}(2),  207--214 (1986). \doi{10.1016/0097-3165(86)90091-9}

\bibitem{lam1983nonexistence}
Lam, C.W.H., Thiel, L., Swiercz, S., McKay, J.: The nonexistence of ovals in a
  projective plane of order 10. Discrete Mathematics  \textbf{45}(2-3),
  319--321 (1983). \doi{10.1016/0012-365X(83)90049-3}

\bibitem{liang2017empirical}
Liang, J.H., {Govind V.K.}, H., Poupart, P., Czarnecki, K., Ganesh, V.: An
  empirical study of branching heuristics through the lens of global learning
  rate. In: International Conference on Theory and Applications of
  Satisfiability Testing. pp. 119--135. Springer (2017).
  \doi{10.1007/978-3-319-66263-3\_8}

\bibitem{macwilliams1973existence}
MacWilliams, F.J., Sloane, N.J.A., Thompson, J.G.: On the existence of a
  projective plane of order 10. Journal of Combinatorial Theory, Series A
  \textbf{14}(1),  66--78 (1973). \doi{10.1016/0097-3165(73)90064-2}

\bibitem{mathon1981partial}
Mathon, R.: The partial geometries $\operatorname{pg}(5, 7, 3)$. Congressus
  Numerantium  \textbf{31},  129--139 (1981)

\bibitem{mckay2014practical}
McKay, B.D., Piperno, A.: Practical graph isomorphism, {II}. Journal of
  Symbolic Computation  \textbf{60},  94--112 (2014).
  \doi{10.1016/j.jsc.2013.09.003}

\bibitem{nadel2012efficient}
Nadel, A., Ryvchin, V.: Efficient {SAT} solving under assumptions. In:
  International Conference on Theory and Applications of Satisfiability
  Testing. pp. 242--255. Springer (2012). \doi{10.1007/978-3-642-31612-8\_19}

\bibitem{read1978every}
Read, R.C.: Every one a winner or how to avoid isomorphism search when
  cataloguing combinatorial configurations. In: Annals of Discrete Mathematics,
  vol.~2, pp. 107--120. Elsevier (1978). \doi{10.1016/S0167-5060(08)70325-X}

\bibitem{roy2011confirmation}
Roy, D.J.: Confirmation of the Non-existence of a Projective Plane of Order 10.
  Master's thesis, Carleton University (2011). \doi{10.22215/etd/2011-09202}

\bibitem{royle1998orderly}
Royle, G.F.: An orderly algorithm and some applications in finite geometry.
  Discrete Mathematics  \textbf{185}(1-3),  105--115 (1998).
  \doi{10.1016/S0012-365X(97)00167-2}

\bibitem{wetzler2014drat}
Wetzler, N., Heule, M.J.H., {Hunt Jr.}, W.A.: {DRAT-trim}: Efficient checking
  and trimming using expressive clausal proofs. In: International Conference on
  Theory and Applications of Satisfiability Testing. pp. 422--429. Springer
  (2014). \doi{10.1007/978-3-319-09284-3\_31}

\bibitem{zulkoski2017combining}
Zulkoski, E., Bright, C., Heinle, A., Kotsireas, I., Czarnecki, K., Ganesh, V.:
  Combining {SAT} solvers with computer algebra systems to verify combinatorial
  conjectures. Journal of Automated Reasoning  \textbf{58}(3),  313--339
  (2017). \doi{10.1007/s10817-016-9396-y}

\end{thebibliography}

\newpage

\section*{Appendix}

Below is a visual depiction of one of the completions of blocks 2--6.  It
includes the twelve columns of the oval and the 45 columns from blocks 2--6:

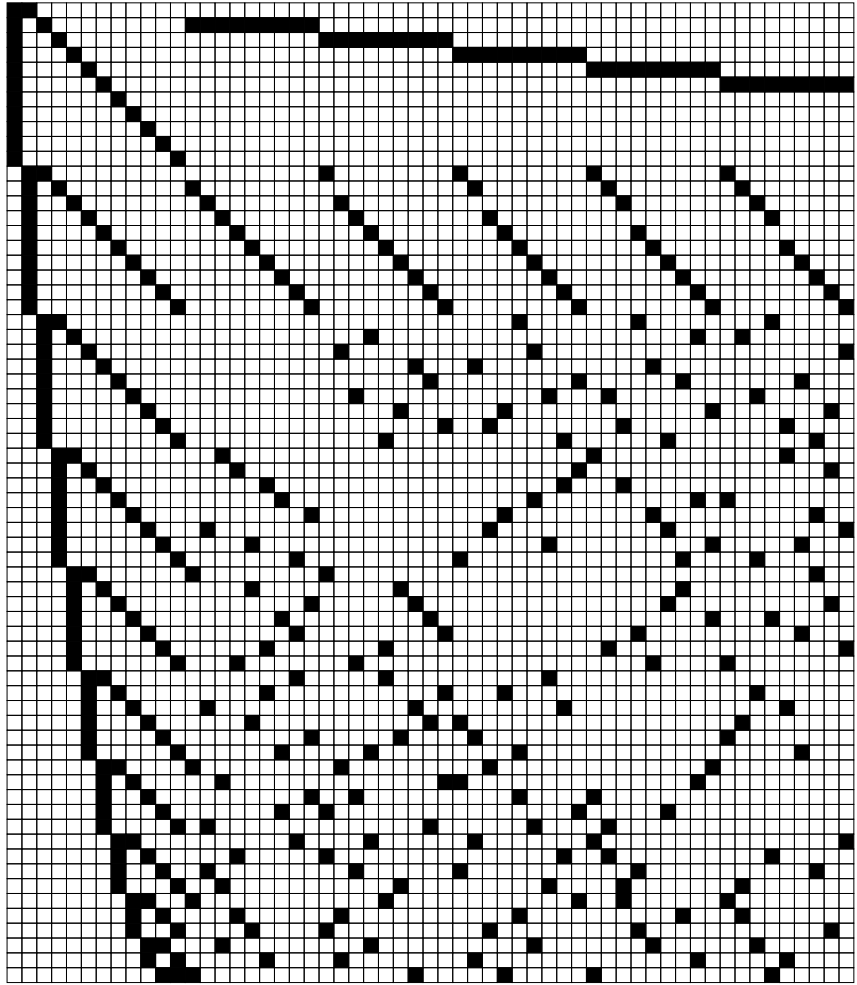
\input{completion.tikz}

\newpage

\noindent Below is a visual flowchart outlining our search implementation:

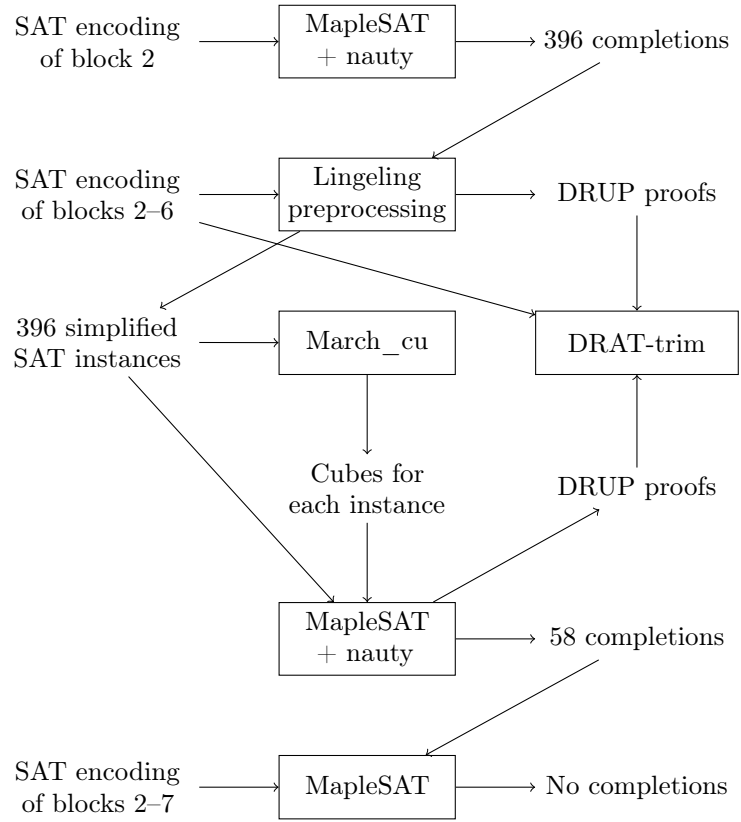
\begin{center}
\begin{tikzpicture}[align=center,node distance=3em]
\node[text width=6em,rectangle,draw](maplesat){MapleSAT + nauty};
\node[left=of maplesat,text width=7em](input){SAT encoding of block 2};
\node[right=of maplesat,text width=7em](sol){396 completions};
\node[below=of maplesat,rectangle,draw,text width=6em](lingeling){Lingeling preprocessing};
\node[left=of lingeling,text width=7em](input2){SAT encoding of blocks 2--6};
\node[right=of lingeling,text width=7em](proofs){DRUP proofs};
\node[below=of input2,text width=7em](input3){396 simplified SAT instances};
\node[below=of lingeling,text width=6em,rectangle,draw,minimum height=2.4em](march){March\_cu};
\node[below=of march,text width=6em](cubes){Cubes for each instance};
\node[below=of cubes,text width=6em,rectangle,draw](maplesat2){MapleSAT + nauty};
\node[right=of march,text width=7em,rectangle,draw,minimum height=2.4em](cat){DRAT-trim};
\node[right=of cubes,text width=7em](proofs2){DRUP proofs};
\node[right=of maplesat2,text width=7em](comp){58 completions};
\node[below=of maplesat2,text width=6em,rectangle,draw,minimum height=2.4em](maplesat3){MapleSAT};
\node[left=of maplesat3,text width=7em](input4){SAT encoding of blocks 2--7};
\node[right=of maplesat3,text width=7em](result){No completions};
\draw[->](input)--(maplesat);
\draw[->](maplesat)--(sol);
\draw[->](sol)--(lingeling);
\draw[->](input2)--(lingeling);
\draw[->](lingeling)--(proofs);
\draw[->](lingeling)--(input3);
\draw[->](input3)--(march);
\draw[->](march)--(cubes);
\draw[->](input3)--(maplesat2);
\draw[->](cubes)--(maplesat2);
\draw[->](maplesat2)--(proofs2);
\draw[->](maplesat2)--(comp);
\draw[->](comp)--(maplesat3);
\draw[->](input4)--(maplesat3);
\draw[->](maplesat3)--(result);
\draw[->](proofs)--(cat);
\draw[->](proofs2)--(cat);
\draw[->](input2)--(cat);
\end{tikzpicture}
\end{center}

\noindent The output of DRAT-trim is a collection of compressed DRUP proofs and
a verification of the SAT-based steps of the primary search.

\end{document}